\documentclass[aps,twocolumn,amssymb,prc,showpacs]{revtex4-1}

\usepackage{graphicx}
\usepackage{multirow}
\usepackage[breaklinks,colorlinks,urlcolor=blue,citecolor=blue]{hyperref}
\usepackage{mathrsfs}
\usepackage{amsmath}
\usepackage{soul}
\usepackage{color}
\usepackage[normalem]{ulem}

\begin{document}

\title{Simultaneous Fitting of Neutron Star Structure and Cooling Data}
\author{Spencer Beloin$^{1}$}
\author{Sophia Han$^{1,2,3}$} 
\author{Andrew W. Steiner$^{1,4}$} 
\author{Khorgolkhuu Odbadrakh$^{5}$} 
\affiliation{$^{1}$Department of Physics and Astronomy, University of
  Tennessee, Knoxville, Tennesee 37996, USA}
\affiliation{$^{2}$Department of Physics and Astronomy, Ohio University,
  Athens, Ohio 45701, USA}
\affiliation{$^{3}$Department of Physics, University of California
  Berkeley, Berkeley, California ~94720, USA}
\affiliation{$^{4}$Physics Division, Oak Ridge National Laboratory, Oak
  Ridge, TN 37831, USA}
\affiliation{$^{5}$Joint Institute for Computational Sciences,
  Oak Ridge National Laboratory, Oak Ridge, Tennesee 37830, USA}
\begin{abstract}
  Using a model for the equation of state and composition of dense
  matter and the magnitude of singlet proton superconductivity and
  triplet neutron superfluidity, we perform the first simultaneous fit
  of neutron star masses and radii determined from observations of
  quiescent low-mass x-ray binaries and luminosities and ages
  determined from observations of isolated neutron stars. We find that
  the Vela pulsar strongly determines the values inferred for the
  superfluid/superconducting gaps and the neutron star radius. We
  find, regardless of whether or not the Vela pulsar is included in
  the analysis, that the threshold density for the direct Urca process
  lies between the central density of 1.7 and 2 solar mass neutron
  stars. We also find that two solar mass stars are unlikely to cool
  principally by the direct Urca process because of the suppression by
  neutron triplet superfluidity.
\end{abstract}


\maketitle

\section{Introduction}

The determination of the equation of state (EOS) of dense matter at
densities beyond the saturation density has been dominated by the
analysis of photon-based observations which lead to determinations of
neutron star masses and
radii~\cite{Cameron59ns,Lattimer01,Steiner10te,Steiner13tn}. The
observation of gravitational waves from the binary neutron star merger
GW170817~\cite{Abbott17go} by the LIGO-Virgo collaboration
significantly changes this picture. Gravitational wave-based probes of
the neutron star tidal deformability have smaller systematic
uncertainties than electromagnetic probes of the neutron star
radius~\cite{Lackey14}. Thus, depending on the mass distribution of
progenitor masses of neutron stars that merge~\cite{Fryer15tf},
gravitational wave observations can potentially drastically decrease
the uncertainties in the EOS in the near future.

However, knowledge of the equation of state alone is insufficient to
describe observed neutron star phenomenology. Neutron star
cooling~\cite{Yakovlev04,Page04mc} and pulsar
glitches~\cite{Piekarewicz14pg} both depend strongly on the role
played by superfluidity and superconductivity~\cite{Page14ss}. On the
other hand, neither neutron star mass and radius observations nor
gravitational wave detections are likely to constrain superfluidity or
superconductivity in the near future. Thus, a joint analysis of data
that constrains the EOS and the nature of pairing in dense matter is
required to refine our understanding of neutron star observations.

In addition, determination of the equation of state does not ensure
knowledge of the composition of dense matter. This was demonstrated in
Ref.~\cite{Alford05hs} where it was found that a mass-radius curve
with neutrons and protons alone was nearly identical to that from a
model containing deconfined quarks. Neutron star cooling, on the other
hand, is sensitive to the neutron-to-proton ratio, since that controls
the threshold for the direct Urca process~\cite{Boguta81ro,Lattimer91du}.

Before this work, technical limitations forced quantitative
analyses of neutron star masses and radii and neutron star thermal
evolution to be almost entirely separate (for an exception, see
Ref.~\cite{Ho15}). In this work, we perform the first combined
analysis of mass and radius data and luminosity and age data for
isolated neutron stars with a well-defined likelihood function. In
addition, we ensure that our model reproduces the structure of heavy
nuclei. Since we allow the mass of each neutron star to vary, we can
go beyond the ``minimal cooling'' model first envisaged in
Ref.~\cite{Page04mc} by varying the EOS and including the direct Urca
process.

\section{Method}

We presume that neutron stars contain only neutrons,
protons, electrons, and muons, leaving the description of exotic
hadrons or deconfined quarks to future work. We employ Bayesian
inference, and all parameters below are taken to have uniform prior
distributions. It is assumed that the data for each neutron star and
all of the nuclear structure data are independent. The conditional
probability (we will refer to this as the likelihood) will thus be a
product of terms that we describe below.

Nuclei and the equation of state are modeled with a covariant
field-theoretical Lagrangian from Ref.~\cite{Steiner05ia}, based on
earlier work in, e.g. Ref.~\cite{Serot86tr}. To be as
flexible as possible with regard to the description of high-density
matter, the model in Ref.~\cite{Steiner05ia} (see Table 3) used 17
parameters. We intentionally employ a relatively large parameter space
to ensure that models are not prematurely ruled out. We compute the
structure of $^{208}\mathrm{Pb}$ and $^{90}\mathrm{Zr}$ in the Hartree
approximation, and thus the first term in the likelihood is a Gaussian
for the binding energy and charge radius of those two nuclei. We
presume 2\% uncertainties for the structure data, larger than the
experimental accuracy, to allow for some systematics due to
limitations in the Hartree approximation, similar to the method used
in Refs.~\cite{Steiner05ia,Steiner13cs} (see also a similar
calculation in Ref.~\cite{Chen14br}). We choose only one medium
mass and one large mass nucleus because our goal is only to ensure
that our nucleon-nucleon interaction is reasonable rather than to
reproduce the properties of all nuclei. Pairing is not included in
the nuclear structure calculations. We constrain the nuclear
incompressibility to be between 220 and 260 MeV~\cite{Garg18tc}. To be
consistent with recent progress in nuclear theory, we restrict the
symmetry energy at the nuclear saturation density, $S$, to be less
than 36 MeV and the slope of the symmetry energy at the nuclear
saturation density, $L$, to be less than 80 MeV~\cite{Lattimer14co}. A
posterior correlation between $S$ and $L$ is automatically obtained
from the fit to the nuclear structure data.

For the neutron star mass and radius data, we use the results obtained
from the seven quiescent low-mass x-ray binary (QLMXB) neutron stars
in the baseline data set from Refs.~\cite{Steiner18ct,Shaw18tr}. The
associated term in the likelihood function is constructed as in, e.g.
Eq.~102 of Ref.~\cite{Lattimer14co} (see also the more generic
formalism developed in Ref.~\cite{Steiner18ta}). A nuisance variable
is required to parametrize the curve, and we use the neutron star
mass for this purpose, giving seven new parameters that are constrained
to be larger than 1 solar mass. Each QLMXB (except for the neutron
star in $\omega$ Cen) can have an atmosphere of either hydrogen or
helium. We construct a new parameter, $\eta$, for each neutron star
ranging between 0 and 1. If $\eta$ is less than 2/3, then a hydrogen
atmosphere is assumed, otherwise a helium atmosphere is
assumed~\cite{Steiner18ct}. The only substantial difference
between the mass and radius input from Ref.~\cite{Steiner18ct} and
this work is the update of the M13 data as reported in
Ref.~\cite{Shaw18tr}.

For the isolated neutron star cooling data, we use the luminosity and
age data from Table I of Ref.~\cite{Beloin18cs}, and the associated
term in the likelihood function is based on Eq. 4 of
Ref.~\cite{Beloin18cs}. This data set does not include neutron
stars with magnetic fields that are estimated to be larger than
$10^{14}$ G, as the surface luminosity is more likely to be impacted
by the magnetic field in these cases. We do not include the neutron
stars in the Cassiopeia A (``Cas A'') and HESS J1731$-$347 supernova
remnants. The thermal evolution of the neutron star in Cas A is still
not well understood (see
Refs.~\cite{Page11rc,Shternin11,Bonanno13tn,Elshamouty13,Posselt13}).
XMMU J173203.3$-$344518 (``J1732'') in HESS J1731$-$347 may be
described by either a carbon atmosphere or a hydrogen
atmosphere~\cite{Klochkov14tn}. Finally, the carbon-atmosphere stars
were shown to not have as strong an impact on the analysis as the Vela
pulsar (PSR B0833$-$45; hereafter ``Vela'') does~\cite{Beloin18cs}.
We also do not include the carbon-atmosphere stars CXOU
J181852.0-150213~\cite{Klochkov16} and CXOU
J160103.1-513353~\cite{Doroshenko18} because pulsations have not
  permitted an age estimate. Below we present results with
and without Vela, due to its strong impact on the posteriors. Each of
the 15 stars requires a mass parameter, and the 9 hydrogen-atmosphere
stars may have a heavy- or light-element envelope, represented by the
variable $\eta$~\cite{Potekhin97} varying between $10^{-17}$ and
$10^{-7}$ (note that this variable $\eta$ from Ref.~\cite{Potekhin97}
refers to the composition of the envelope and differs from the
variable $\eta$ referred to in the preceding paragraph from
Ref.~\cite{Steiner18ct} which refers to the atmosphere composition).
We assume spherically symmetric surface temperatures and that the
magnetic field is small enough to be irrelevant for the cooling.

All theoretical calculations of the singlet proton superconducting gap
and the neutron triplet superfluid gap in the neutron star core
contain uncontrolled approximations. Thus we parametrize the
associated critical temperatures and allow the data to select the
correct values, as in Ref.~\cite{Beloin18cs}. Each gap is described by
a Gaussian function of the Fermi momenta, with height, centroid and
width parameters denoted $T_{C,i}$, $k_{F,\mathrm{peak},i}$ and
$\Delta k_{F,i}$, respectively. To avoid double-counting, for both
neutrons and protons, we constrain $k_{F,\mathrm{peak}}$ to be larger
than the value of $k_F$ at the crust-core transition, and both
$k_{F,\mathrm{peak}}$ and $\Delta k_F$ to be smaller than the value of
$k_F$ in the core of the maximum-mass star. Note that we do not ensure
that the superfluid or superconducting gaps are consistent with the
Lagrangian used to describe the EOS of nucleonic matter. This kind of
consistency, while sometimes important in nuclear
structure~\cite{Chen14rm}, is unnecessary in this context, since the
pairing interaction does not strongly impact the equation of state and
our gaps will always be less than 1 MeV, which is small compared to
the nucleon Fermi energy. We include the direct Urca process, the
modified Urca process, the pair-breaking neutrino emissivity, and
other cooling processes enumerated in Refs.~\cite{Page04mc,Page09ne}.

To account for the fact that our Lagrangian may not fully
describe nucleons in the neutron star core, we add a two-parameter
polytrope to our EOS following Ref.~\cite{Han17co}, which begins at
twice the nuclear saturation density and either softens or stiffens
the EOS depending on the sign of the coefficient and the magnitude of
the exponent. We automatically reject any models that are acausal
(because of the additional polytrope), have a maximum mass less than 2
$\mathrm{M}_{\odot}$, or for which any neutron stars have a mass
larger than the maximum mass of the EOS. 

Our model has a total of 64 parameters, and traditional Bayesian
inference applied to this problem requires the solution to the TOV
equations, the nuclear structure calculations, and several cooling
curves (to handle the variation in mass and envelope composition) at
each point. To make it computationally tractable, we do not fully
compute every point. Instead, we construct a library of exact
calculations using a simple MCMC method. We fit the likelihood
function to a 64-dimensional Gaussian (including covariances between
all of the parameters) and then directly sample that Gaussian to
  generate parameter sets. To predict quantities other than the
likelihood (cooling curves, $M-R$ curves, gaps, etc.) from the library
using samples from the Gaussian, we use inverse-distance-weighted
interpolation. Sampling the emulator sometimes results in
  points which give unphysical values for the maximum mass
  or the speed of sound, so these points are removed to obtain the
  final results.

The full likelihood function is thus
\begin{equation}
  {\cal L} = {\cal L}_{\mathrm{nuclei}}
  {\cal L}_{\mathrm{QLMXB}} {\cal L}_{\mathrm{INS}} \, .
  \label{eq:like}
\end{equation}
The first term in Eq.~\ref{eq:like} is the
contribution from the nuclear
structure data,
\begin{eqnarray}
  {\cal L}_{\mathrm{nuclei}} &=&
  \exp \left[ -\frac{\left(E_{\mathrm{Zr}} -
      \tilde{E}_{\mathrm{Zr}}\right)^2}{
  \left(0.02 \tilde{E}_{\mathrm{Zr}}\right)^2} \right]
  \exp \left[ -\frac{\left(R_{\mathrm{Zr}} -
      \tilde{R}_{\mathrm{Zr}}\right)^2}{
      \left(0.02 \tilde{R}_{\mathrm{Zr}}\right)^2} \right] \nonumber \\
  && \exp \left[ -\frac{\left(E_{\mathrm{Pb}} -
        \tilde{E}_{\mathrm{Pb}}\right)^2}{
      \left(0.02 \tilde{E}_{\mathrm{Pb}}\right)^2} \right]
  \exp \left[ -\frac{\left(R_{\mathrm{Pb}} -
      \tilde{R}_{\mathrm{Pb}}\right)^2}{
      \left(0.02 \tilde{R}_{\mathrm{Pb}}\right)^2} \right] \, ,
  \nonumber \\
\end{eqnarray}
where $E_{i}$ is the calculated binding energy of nucleus $i$,
$\tilde{E}_i$ is the experimental binding energy, $R_i$, is the
calculated charge radius, and $\tilde{R}_i$ is the experimental
charge radius. The second term in Eq.~\ref{eq:like}
comes from the QLMXB observations
\begin{eqnarray}
  {\cal L}_{\mathrm{QLMXB}} &=&
  \prod_{i=1}^{7}
  \left\{
  \theta(2/3-\eta_i)
        {\cal D}_{i,\mathrm{H}}[R(M_i),M_i] +\right. \nonumber \\
        &&
        \left.
        \theta(\eta_i-2/3)
              {\cal D}_{i,\mathrm{He}}[R(M_i),M_i] \right\}
              \label{eq:QLMXB}
\end{eqnarray}
where $i$ runs over the seven QLMXBs, $\eta_i$ is the atmosphere parameter
described above, $M_i$ is the neutron star mass, $R(M_i)$ is the
radius as a function of mass determined by the TOV equations, ${\cal
  D}_{i,\mathrm{H}}$ is the distance-uncertainty averaged probability
distribution obtained from the x-ray observations assuming a hydrogen
atmosphere, and ${\cal D}_{i,\mathrm{He}}$ is the probability
distribution assuming a helium atmosphere. The third term in
Eq.~\ref{eq:like} comes from the isolated neutron star (INS) cooling
observations
\begin{eqnarray}
  {\cal L}_{\mathrm{INS}} &=&
  \prod_{j=1}^{16}{\sum_{k}{\sqrt{\left\{\left[\frac{d\widehat{L}
            (\eta_j,M_j)}
          {d\widehat{t}}\right]_k^{2}+1\right\}}}} \nonumber \\
  && \times \exp\left\{\frac{-\left[\widehat{t}_k-\widehat{t}_{j}\right]^{2}}
       {2(\delta \widehat{t}_j)^{2}}\right\} 
       \nonumber \\
       && \times\exp\left\{\frac{-\left[\widehat{L}_k(\eta_j,M_j)-
           \widehat{L}_{j}\right]^{2}}
                 {2(\delta \widehat{L}_j)^{2}}\right\} \, .
\end{eqnarray}
where $j$ is an index over the 16 isolated neutron stars
(or 15 if Vela is not included),
$\widehat{t}=\mathrm{log}_{10}[t/(10^2~\mathrm{yr})]/5$ is
a logarithmic coordinate for the time,
$\widehat{L}=\mathrm{log}_{10}[L/(10^{30}~\mathrm{erg/s})]/4$
is a logarithmic coordinate for the luminosity, $k$ is an
index over a uniform grid in $\widehat{t}$ (see Ref.~\cite{Beloin18cs}
for details), $\eta_j$ is a parameter for
the envelope composition which runs from $10^{-19}$ to $10^{-7}$,
$\widehat{L}_k(\eta_j,M_j)$ is the theoretical luminosity for neutron star
$j$ at grid point $k$ (which depends on envelope composition and
mass), $\widehat{L}_j \pm \delta \widehat{L}_j$ is the observed luminosity
with its associated uncertainty, $\widehat{t}_j \pm \delta \widehat{t}_j$
is the observed age with its associated uncertainty, and the term
under the square root is a geometric factor (see Ref.~\cite{Steiner18ta}).
The computation of $\widehat{L}_k(\eta_j,M_j)$ requires first a
solution of the TOV equations to determine the structure, and then
a solution of the stellar evolution equations to determine the
cooling.

\section{Results}

\begin{figure}
  \includegraphics[width=0.23\textwidth]{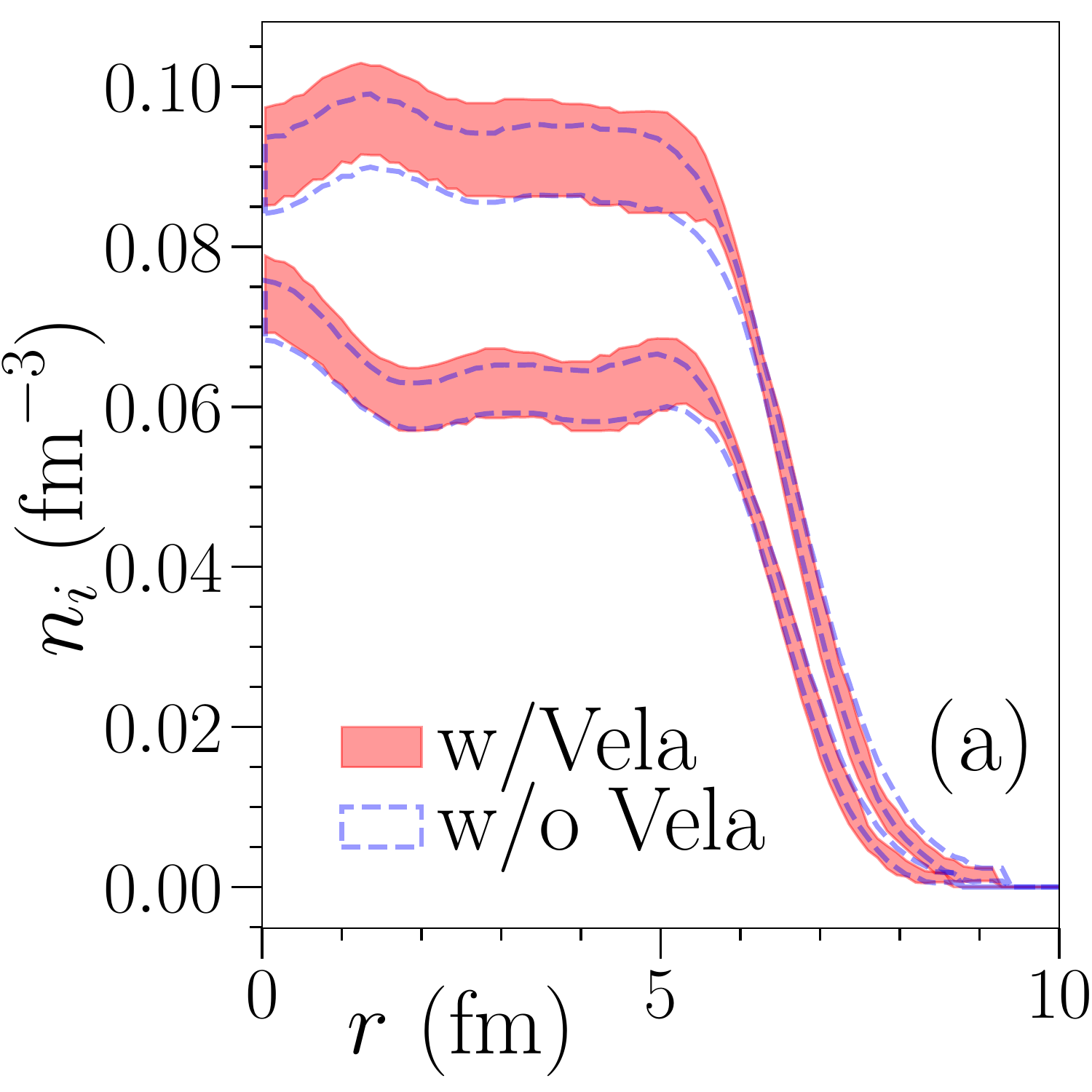}
  \includegraphics[width=0.23\textwidth]{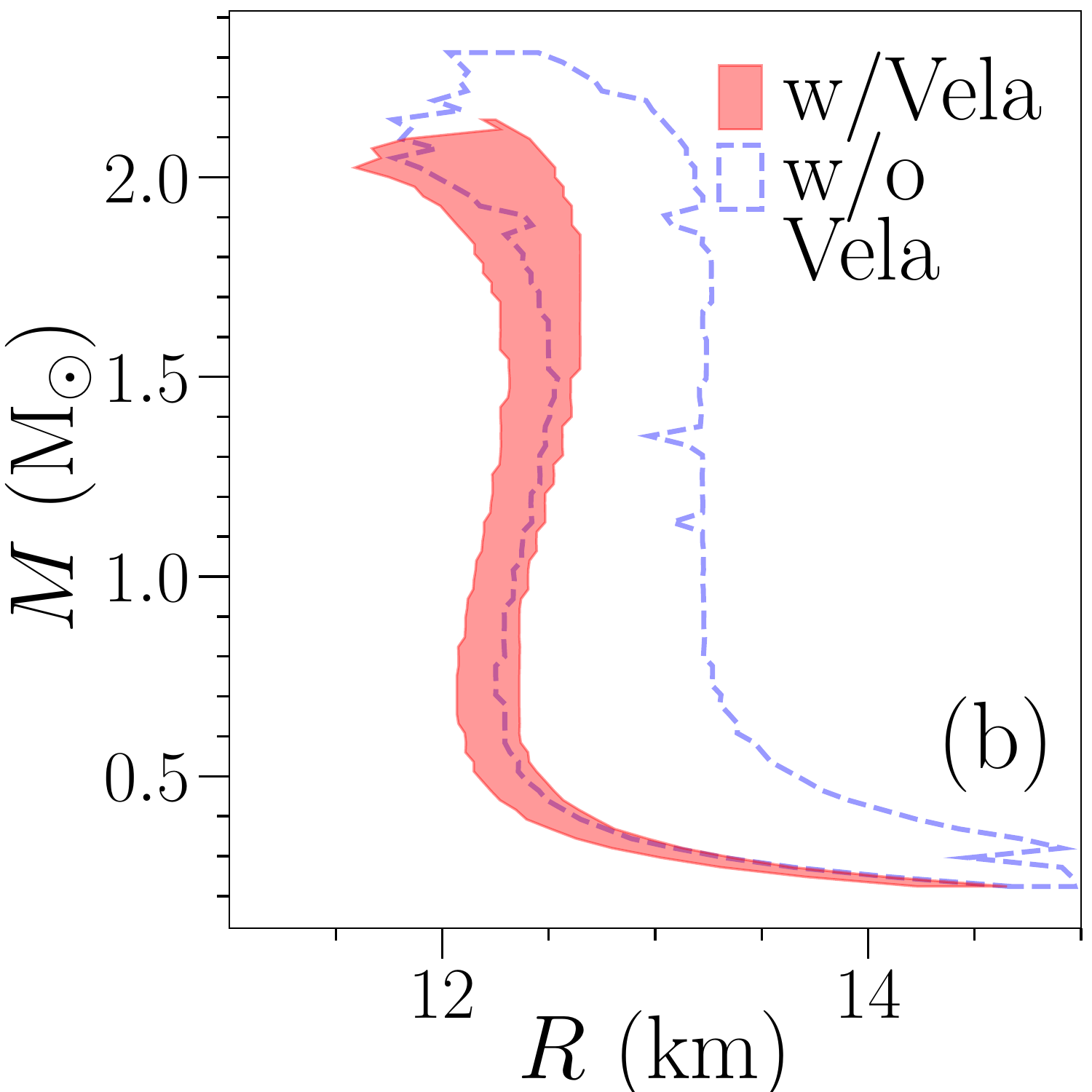}
  \includegraphics[width=0.23\textwidth]{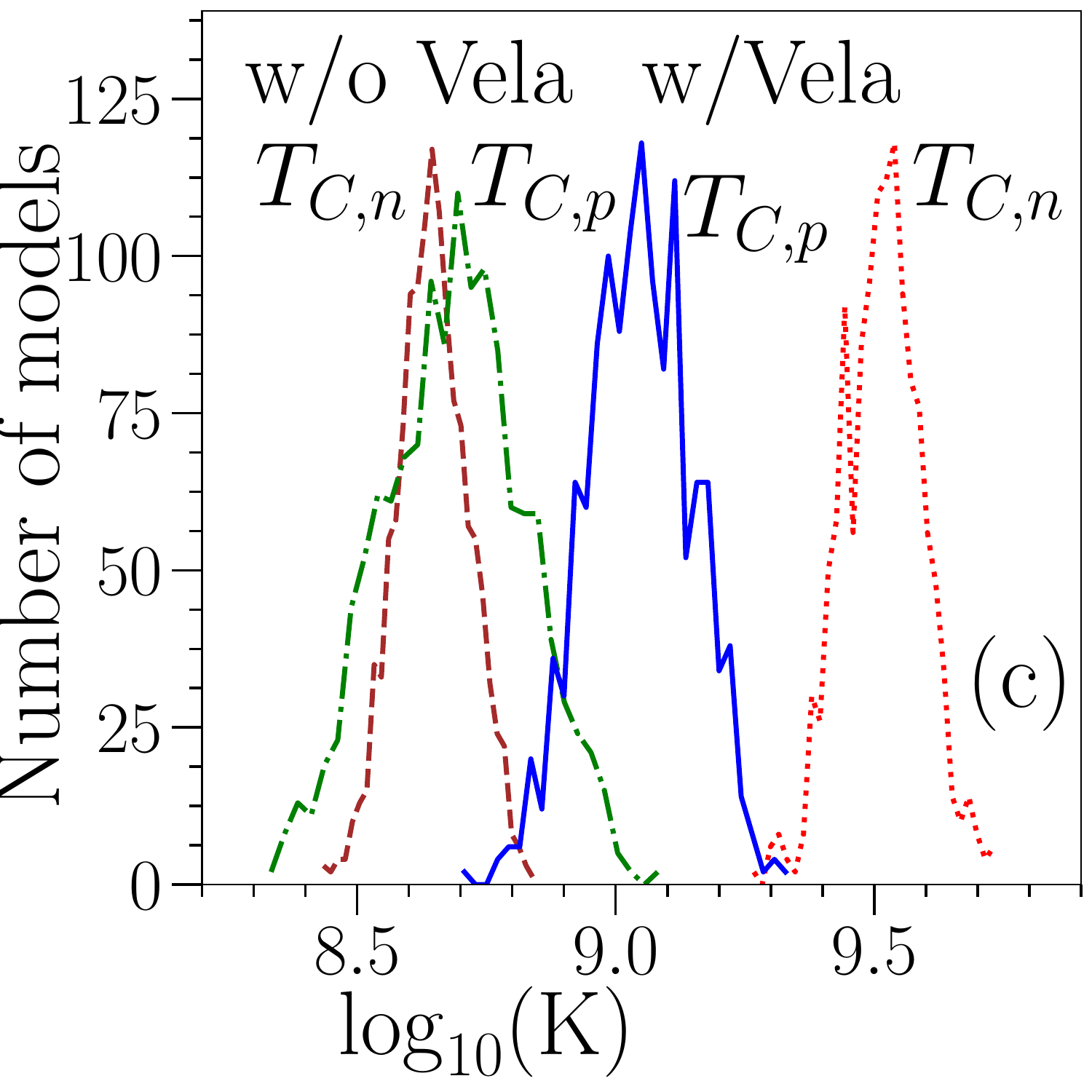}
  \includegraphics[width=0.23\textwidth]{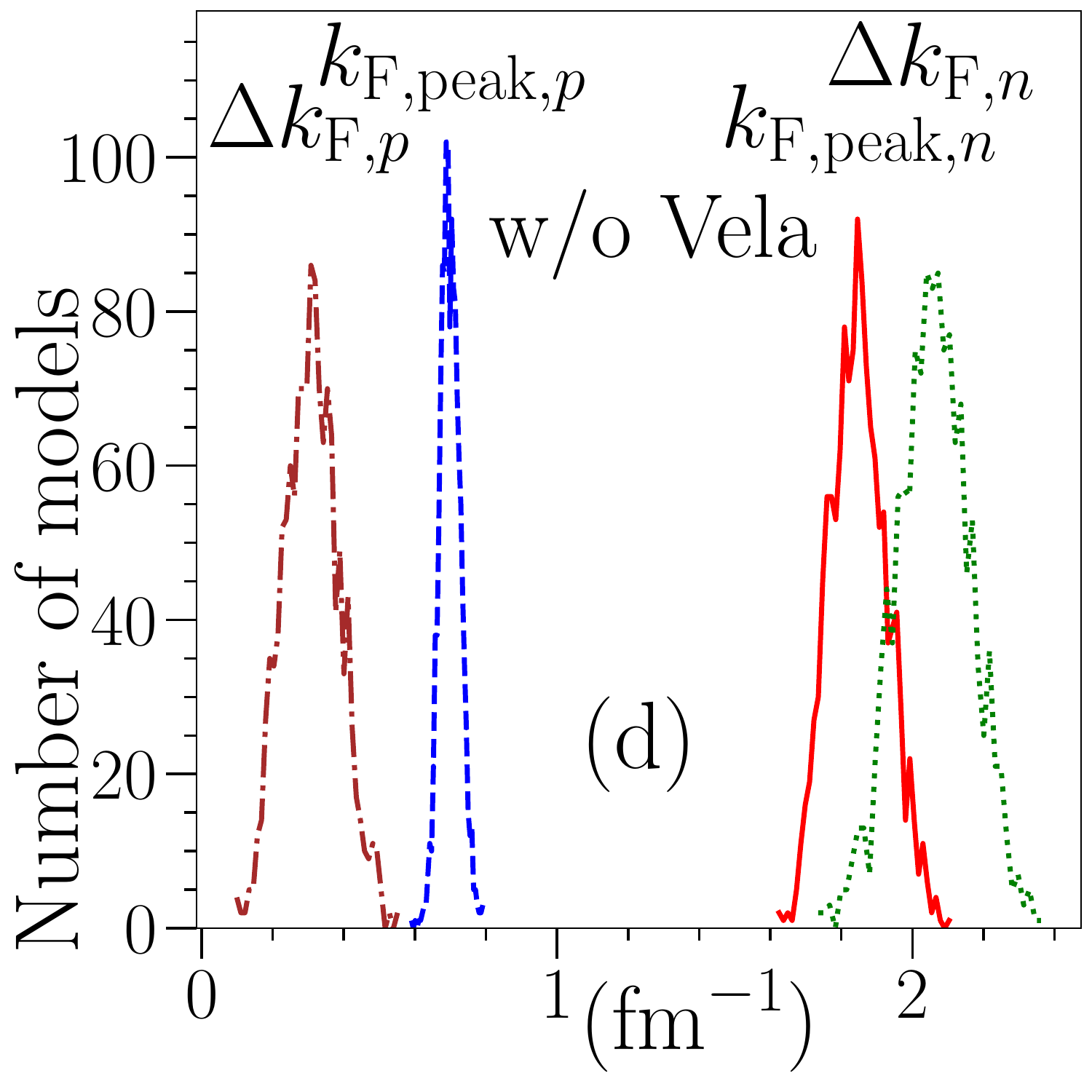}
  \includegraphics[width=0.23\textwidth]{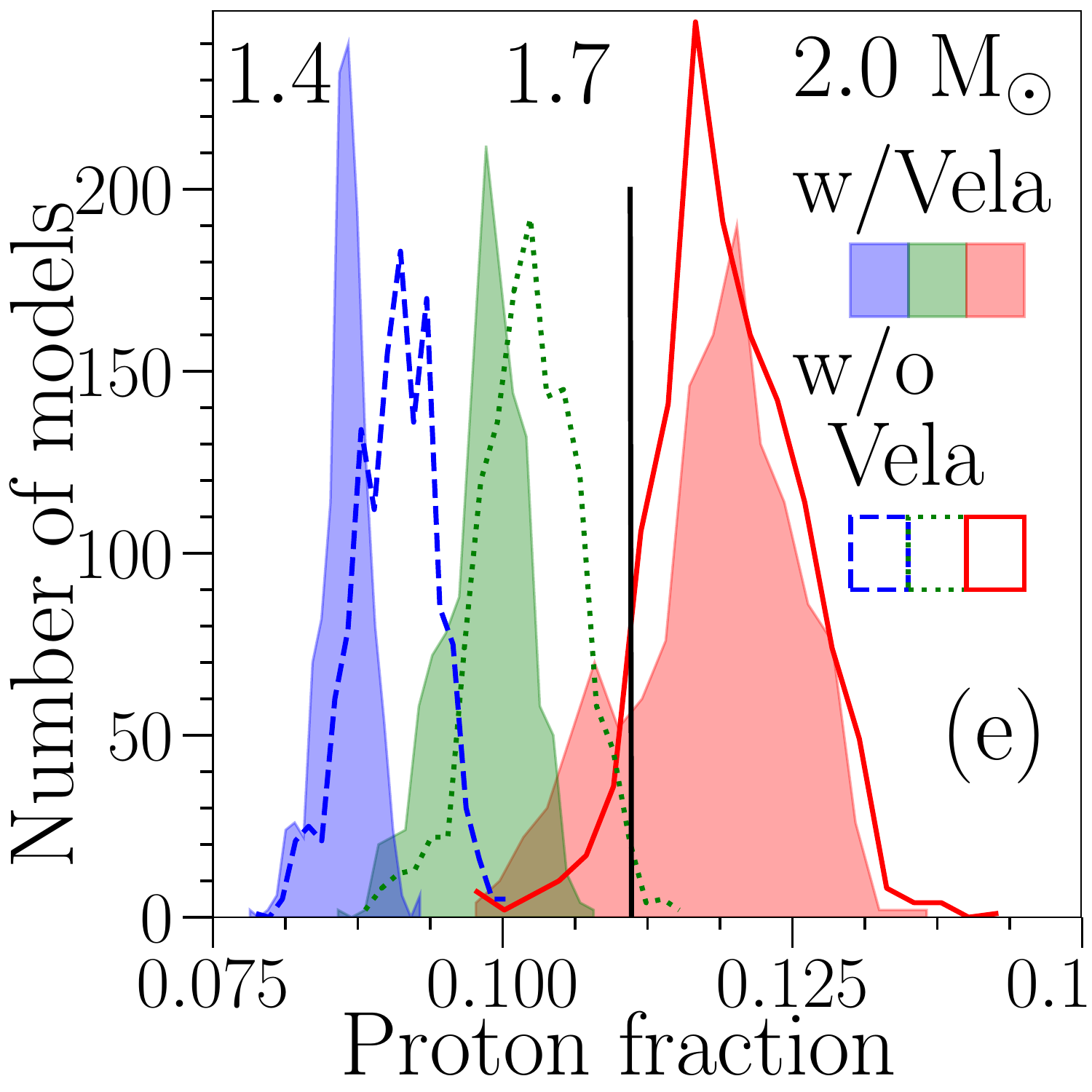}
  \includegraphics[width=0.23\textwidth]{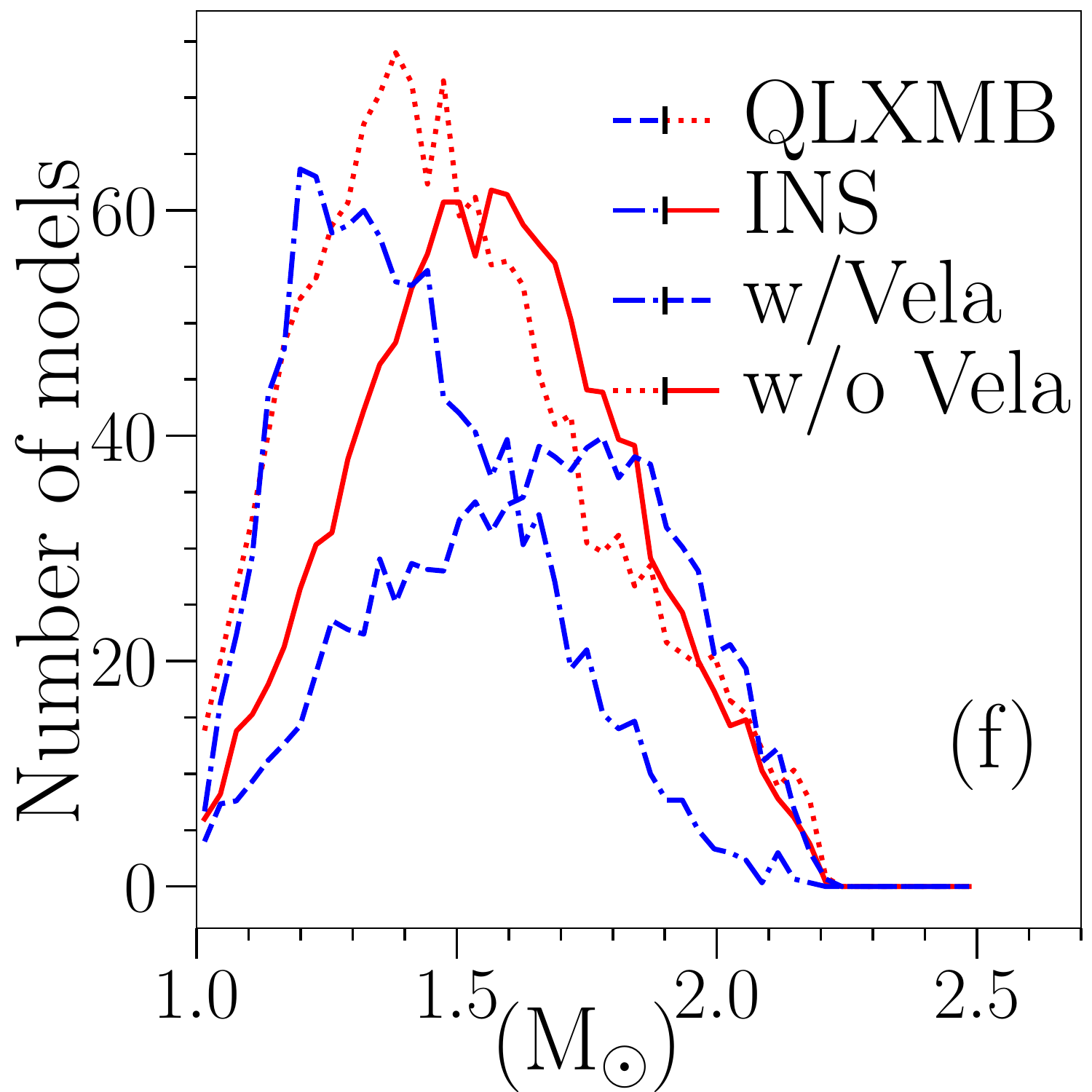}
  \caption{All panels except (d) show results with and without Vela.
    For display purposes, simulations with and without Vela are fixed
    to have the same overall normalization. Panel (a): 95\% credible
    intervals for the neutron and proton density profiles for
    $^{208}$Pb. Panel (b): 95\% credible intervals for the mass-radius
    curves. Panel (c): Histograms for the maximum value of the proton
    singlet and neutron triplet gaps. Panel (d): Histograms for the
    other two gap parameters for neutrons and protons for the analysis
    excluding Vela. Panel (e): Histograms for the central proton
    fraction of 1.4, 1.7, or 2.0~$\mathrm{M}_{\odot}$ neutron stars.
    Panel (f): Posterior mass distributions for the seven QLMXB
    neutron stars or the 14 isolated cooling neutron stars (or 15
    stars when Vela is included).}
  \label{fig:1}
\end{figure}

The posterior distributions for the neutron and proton density
distributions are given in Fig.~\ref{fig:1}(a). Regardless
of whether or not Vela is included, the charge radii are within 2\% of
the experimental value of 5.5 fm. (As found in Ref.~\cite{Beloin18cs},
Vela has a strong impact on the results because its luminosity is
comparatively small given its estimated age.) Our 95\% credible range
for the neutron skin thickness of $^{208}\mathrm{Pb}$ is 0.15 to 0.20
fm (0.143 to 0.157 fm) without (with) Vela. The addition of Vela
pushes several parameters, including those which control the
nucleon-nucleon interaction, to extreme values. When Vela is included,
the value of $L$ (and thus the value of the skin thickness) is pushed
towards smaller values to decrease the specific heat and
increase the cooling. Note that when $L$ is smaller, the isospin
asymmetry in the central region increases to ensure that the
total neutron and proton number is fixed~\cite{Steiner05ia}.

The distribution for mass-radius curves is plotted in 
Fig.~\ref{fig:1}(b). We find the radius of a 1.4 solar mass star to be
between 12.4 and 13.1 km (12.35 to 12.65 km) without (with) Vela. When
Vela is included, the smaller value of $L$ leads to smaller neutron
star radii. These ranges are consistent with that obtained from QLMXBs
alone (11.0 to 14.3 km)~\cite{Steiner18ct}, and also with the radius
inferred by the GW170817 merger (9.1 to 12.8 km)~\cite{Abbott18mo}.
There is some jitter in the radius limits due to small statistics.

The peak values of critical temperatures for the proton and neutron
superfluid are presented in Fig.~\ref{fig:1}(c). As in
Ref.~\cite{Beloin18cs}, we find significantly different results
depending on whether or not Vela was included: the gaps are
significantly larger when Vela is included, which helps maximize the
neutrino emissivity from the breaking and formation of Cooper
pairs~\cite{Flowers76a,Gusakov04,Leinson06ne,Leinson06vc,Kolomeitsev08ne,Steiner09sr,Page09ne}.
(Note that, as shown in Fig. 6 of Ref.~\cite{Page09ne}, the dependence
of the cooling curve on the critical temperature is not monotonic.)
Fig \ref{fig:1} (d) shows the centroid and width parameters for the gaps without
Vela. We find that $\Delta k_{F,n}$ is large with or without Vela,
which effectively prevents the direct Urca process from occurring in
the core. This is consistent with the observational data, as no
confirmed isolated neutron stars have very low luminosities (see
Figs.~\ref{fig:2} and \ref{fig:3} below).

Figure \ref{fig:1} (e) in Fig.~\ref{fig:1} displays histograms for the proton
fraction in the cores of 1.4, 1.7, and 2.0~$\mathrm{M}_{\odot}$
neutron stars along with a vertical line at 11\%, the threshold for
non-superfluid neutrons and protons to participate in the direct Urca
process when muons are ignored. We find it unlikely that
1.7~$\mathrm{M}_{\odot}$ neutron stars have proton fractions larger
than 11\%, and very likely that all 2.0~$\mathrm{M}_{\odot}$ stars do.
Adding muons tends to decrease the critical
density~\cite{Lattimer91du} for the Urca process to satisfy energy and
momentum conservation. However, the neutron superfluid quenches the
direct Urca process, even in 2.0~$\mathrm{M}_{\odot}$ stars.

Panel (f) in Fig.~\ref{fig:1} shows the mass posterior for the QLMXB
and isolated neutron star (INS) data sets with Vela (dashed and
dot-dashed lines) and without Vela (dotted and solid lines). Including
Vela decreases the radius of all neutron stars (for fixed
gravitational mass), as shown above. Since the QLMXB observations tend
to fix $R_{\infty}=R/\sqrt{1-2 G M/(R c^2)}$, decreasing $R$ tends to
increase $M$. When Vela is not included, probability distributions for
both QLMXBs and INSs peak near 1.4~$\mathrm{M}_{\odot}$, similar to
that observed in NS-NS and NS-WD binaries with accurate mass
measurements~\cite{Lattimer12}. The mass distributions here imply that
QLMXBs are less massive than INSs when Vela is not included, backwards
from what is expected since QLMXBs gain mass through accretion (see,
e.g. Refs.~\cite{Kiziltan13,Ozel16}). However, the individual masses
have large uncertainties so the precise locations of the peaks in Fig.
1 are not significant (see Appendix for details). More data will be
required to definitively predict mass distributions in this way.

In Fig.~\ref{fig:2}, we show the luminosity-age cooling curves
excluding Vela. The bands refer to the 95\% credible regions for the
luminosity as a function of age. Note that while the gravitational
mass is presumed to be constant over the time period shown, the amount
of light elements in the envelope, parameterized by $\eta$, may
decrease over time as nuclear reactions fuse these light elements into
heavier ones. Cooling curves with Vela included are plotted in
Fig.~\ref{fig:3}. The uncertainty ranges are much smaller,
demonstrating that a much smaller set of parameters is able to
accommodate Vela's small luminosity given its age. The larger
mass stars cool slightly more slowly, as explained in
Ref.~\cite{Page04mc} (see Fig.~25). As would be expected, our
predicted mass for Vela is smaller, $1.25\pm0.25~\mathrm{M}_{\odot}$
to ensure a smaller luminosity for this object. 

\begin{figure}
  \includegraphics[width=0.45\textwidth]{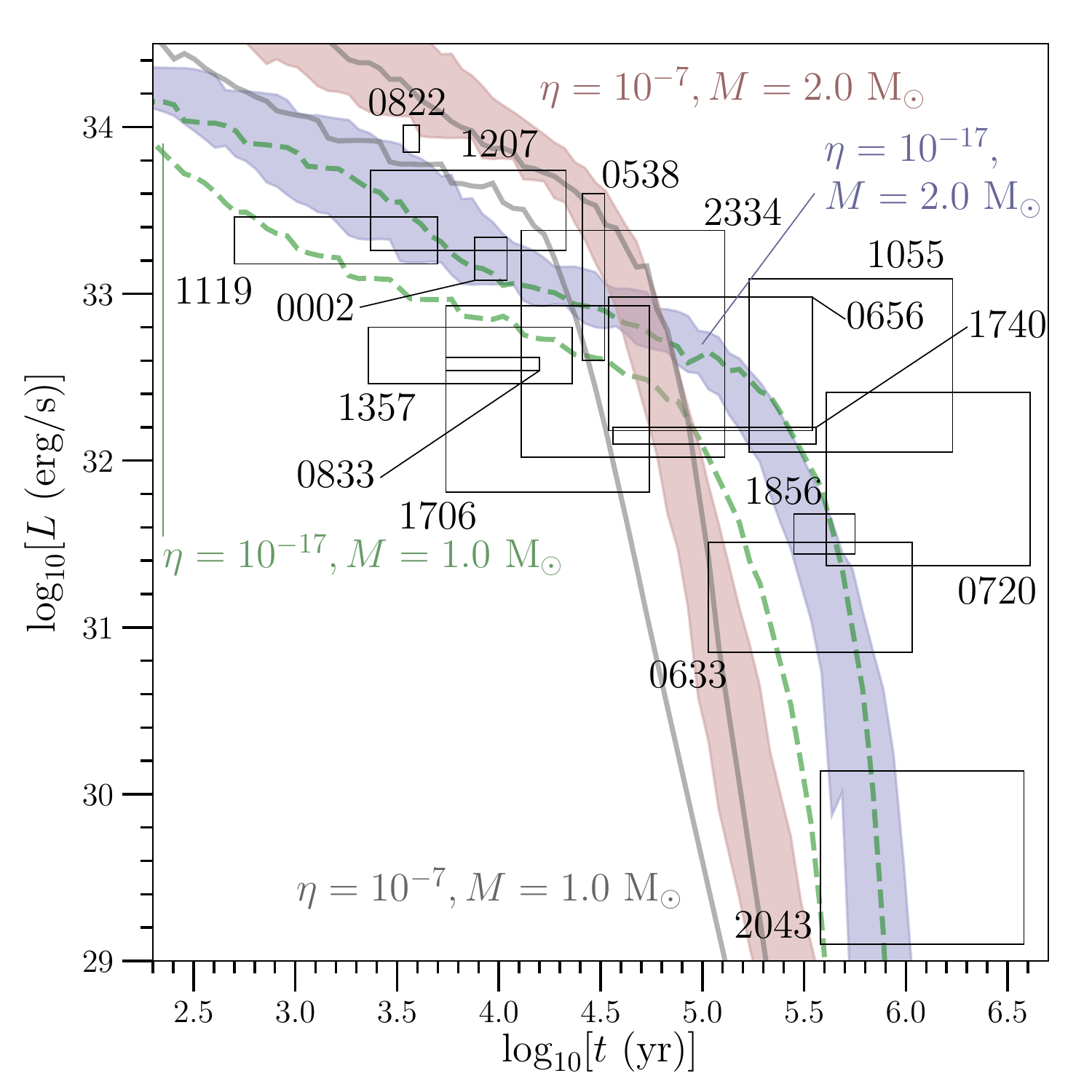}
  \caption{The 95\% credible intervals for the luminosity-age curves
    for different neutron star masses and different amounts of light
    elements in the neutron star envelope (parameterized by $\eta$),
    along with the luminosity and age values obtained from
    observations which were used in this work. Vela (labeled ``0833'')
    is not included in the model results but the data point is
    included in the figure for comparison. }
  \label{fig:2}
\end{figure}

\begin{figure}
  \includegraphics[width=0.45\textwidth]{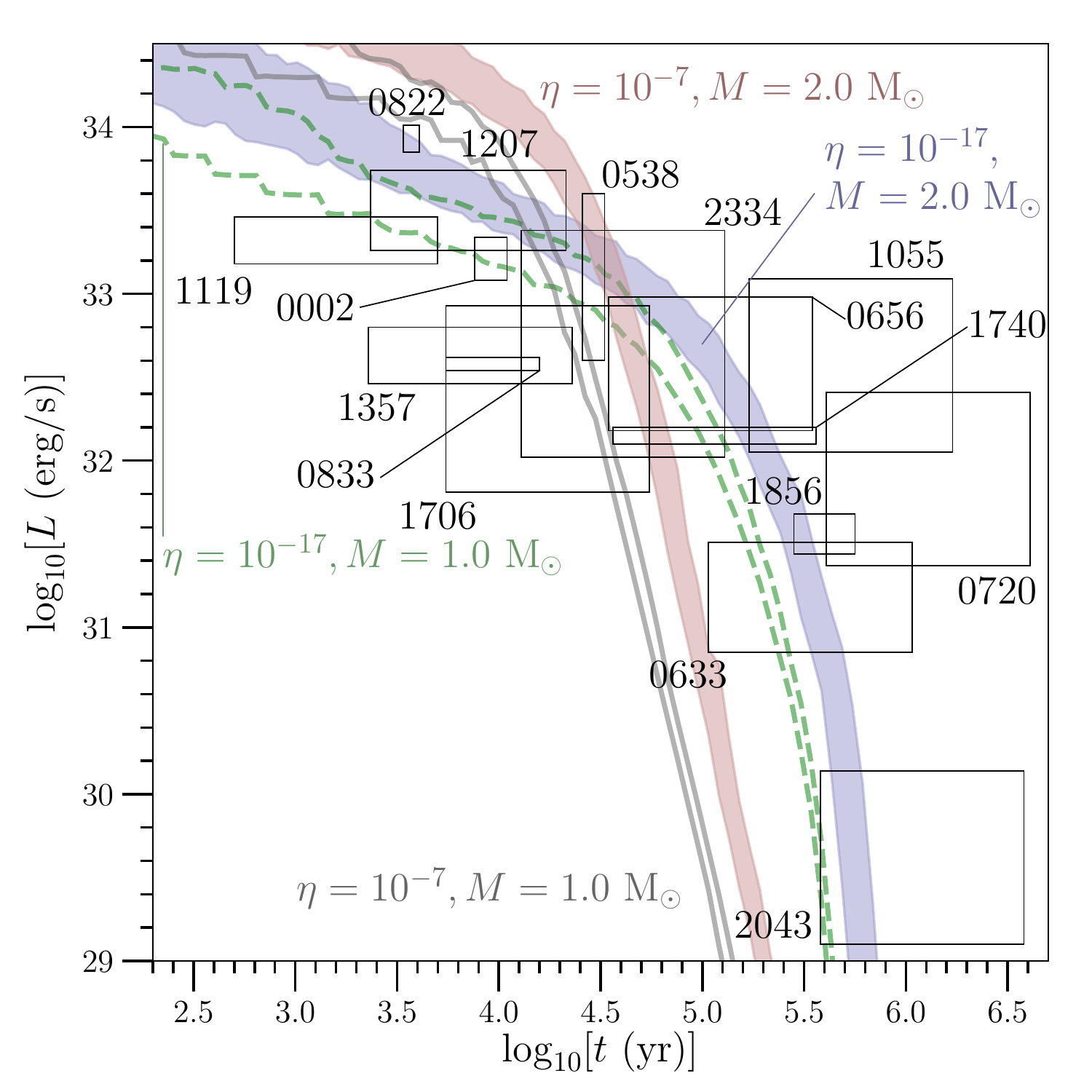}
  \caption{The 95\% credible intervals for the luminosity-age curves
    for different neutron star masses and different amounts of light
    elements in the neutron star envelope (parameterized by $\eta$),
    along with the luminosity and age values obtained from
    observations which were used in this work. Vela (labeled ``0833'')
    is included in the model results.}
  \label{fig:3}
\end{figure}

There are several other recent works describing the cooling of
isolated neutron stars which analyze a smaller data set, and employ a
smaller set of nucleon pairing models with a smaller set of EOS
models, but often take into account exotic degrees of freedom.
The authors of Ref.~\cite{Raduta18co} studied
stars containing hyperons and found a
significant contribution from hyperonic direct Urca processes.
The authors of Ref.~\cite{Negreiros18co} also studied
stars containing hyperons and
found  small neutron star radii and large masses were preferred.
The authors of Ref.~\cite{Potekhin18mn} focused on magnetic fields,
which we do not
include, and were able to include magnetars in their analysis.
The authors of Ref.~\cite{Ofengeim17ad}
included a helpful analytical description of
isolated neutron star cooling.
The authors of Ref.~\cite{Beznogov18co} used a method
similar to ours to analyze the cooling of J1732, and included the
possibility of axion cooling. They applied a single EOS and found large
proton and small neutron gaps.

\section{Conclusions and Discussion}

In this work, we analyzed the implications that our data set, which
includes neutron star masses, radii and luminosity and age
determinations for isolated cooling neutron stars, have on dense
matter. These data suggest that the direct Urca process is unlikely to
play a prominent role in neutron star cooling because of suppression
by neutron triplet superfluidity throughout the core. The proton
fraction in dense matter is unlikely to remain below the threshold for
energy and momentum conservation in the direct Urca process in the
core of more massive neutron stars. The minimal cooling
model~\cite{Page04mc} assumes that the direct Urca process does not
occur, presuming that either the EOS does not allow the direct Urca
process at any density inside neutron stars or that the mass of
isolated neutron stars is not large enough to create a large central
proton density. The results of this work suggest that superfluidity,
rather than the EOS, suppresses the direct Urca process for low to
moderate mass neutron stars.

Many-body correlations (see, e.g.,
Ref.~\cite{Blaschke95,Schaab96,Shternin18}) at high densities can
impact these neutrino processes and can impact our conclusions.
However, it is difficult to include all possible models of these
correlations and the additional uncertainty is difficult to quantify
because the associated many-body calculations are not systematically
improvable. We hope to include this in future work.

In this work we used a field-theoretical Lagrangian, but
performing a similar analysis using a Skyrme model might also be
instructive. However, Skyrme models are also likely to become acausal
at densities below the central density of the maximum mass
star~\cite{Dutra12si}. As also found for mass and radius data alone in
Refs.~\cite{Steiner13tn,Steiner16ns}, we do expect our results to be
sensitive to our prior choices, particularly with regard to the
equation of state. We added an additional polytrope at high densities
and the impact of the polytrope is estimated in the Appendix.

After completing this work we found that updated distance measurements
in Ref.~\cite{Verbiest12} will modify the luminosity inferred for
B1706-44 which was reported in Ref.~\cite{Beloin18cs}. However, this
modification is unlikely to modify our basic conclusions since the
uncertainty in the luminosity for this source is large.

We find that the most likely models selected by this inference appear
incompatible with the cooling of the coldest transiently-accreting
sources, such as SAX J1808.4$-$3658~\cite{Heinke07,Han17co}. The
suppression of the direct Urca emissivity due to pairing found here
cannot explain the low temperatures observed. However, this
incompatibility may be resolved simply if all of the neutron
stars in this work have smaller masses and if SAX J1808 has a large
mass. The best way forward is to simultaneously fit both isolated and
accreting neutron stars, in addition to the information from neutron
star mass and radius determinations. Initial progress on simultaneous
fitting of these two data sets has been made in
Refs.~\cite{Beznogov15a,Beznogov15b} using a handful of EOSs.

The authors of Refs.~\cite{Ofengeim17ad,Ofengeim18} also examined
the cooling of the
Vela pulsar and similarly observed that it was cooling faster than most
of the rest of the population. In contrast to our work, they used the
x-ray spectrum directly to determine its mass and radius. They also
found a relatively low mass for Vela, and they also found that it can be
explained in the context of the minimal cooling model where the direct
Urca process is ignored.

\begin{acknowledgments} 

  We thank Dany Page for publicly releasing his NSCool neutron star
  cooling code. We also thank Craig Heinke and Joonas N\"{a}ttil\"{a}
  for helpful discussions. This work was supported by NSF Grant No. PHY
  1554876 and by the U.S. DOE Office of Nuclear Physics. S.H. is also
  supported by Chandra Grant No. TM8-19002X, NSF Grant No. PHY-1630782, and the
  Heising-Simons Foundation, Grant No. 2017-228. This work used the Extreme
  Science and Engineering Discovery Environment (XSEDE) allocation
  PHY170048 supported by NSF Grant No. ACI-1548562 and
  computational resources from the University of Tennessee and Oak Ridge
  National Laboratory's Joint Institute for Computational Sciences.

\end{acknowledgments} 

\bibliographystyle{apsrev}
\bibliography{paper}

\section*{Appendix: Fit analysis}

This data analysis problem consists of 22 (or 23 when Vela is
included) neutron star observations and 4 nuclear structure data
points, two constraints on the symmetry energy, one constraint on the
incompressibility, and 64 parameters, and thus is formally
underconstrained. Part of the reason the problem is underconstrained
is the additional mass (nuisance) parameter required for each neutron
star to map the location on the curve to the two-dimensional data
point (as shown in Ref.~\cite{Steiner18ta}). However, not all of the
parameters impact the fit equally. For the cooling neutron stars, the
mass and the variable $\eta$ which represents the composition of the
envelope make only a small correction to the cooling curve (unless the
direct Urca process is allowed, which we find is unlikely). In the
midst of these complications, although we do not have the
computational time to explicitly verify this, our problem is
sufficiently underconstrained that we expect our prior choice will
impact our results, as found for QLMXBs and other neutron star
observations in Ref.~\cite{Steiner13tn}.

One of the biggest difficulties with a large parameter space is
  the possibility of overfitting. A complete analysis of overfitting
  requires a much more computationally intensive cross-validation
  which we cannot yet perform. One of the symptoms of overfitting is
  that the parameters are allowed to vary so widely that the model
  becomes physically unrealistic. In lieu of a full cross-validation,
  we discuss the posterior distributions of the parameters and show
  that they are within the expected range and that they do not have
  unreasonably small uncertainties. We find no reason to believe that
  overfitting is impacting the principal results in this paper, as
  described below. 

The superfluid/superconducting critical temperatures which we
  obtain when Vela is not included closely match what was obtained
  previously in Ref.~\cite{Page04mc}, and the fact that Vela has a
  particularly weak luminosity has also been observed
  elsewhere~\cite{Ofengeim17ad,Ofengeim18}, thus it is unsurprising
  that our critical temperatures are modified when this neutron star
  is included. The other superfluid parameters are not frequently
  constrained in the literature, except in Ref.~\cite{Ho15}, which
  analyzed Cas A, so there is little guidance except from uncontrolled
  theory estimates. The posterior values for the envelope
  composition, as described in Ref.~\cite{Beloin18cs}, are also
  reasonable given the relative location of the data points to the
  posterior cooling curves.

All of the posterior values for the neutron star masses are
  between 1.0 and 2.0 solar masses, as would be expected from the
  recent status of simulations of core-collapse
  supernovae~\cite{Hix14ei,Burrows13cp}. All of the posterior mass
  distributions also have large uncertainties, betwen 0.15 and 0.35
  solar masses. When Vela is not included, the lower-right panel
  implies that QLMXBs are about 0.1 $\mathrm{M}_{\odot}$ less massive
  that the isolated neutron stars, but the uncertainty in the
  individual masses means that this is probably not significant. When
  Vela is included, the ordering is reversed, and QLMXBs are more
  massive, as might be expected from evolutionary considerations.

To assess the values of the interaction couplings, we
  compare them with the range of values covered by the
  SFHo~\cite{Steiner13cs}, SFHx~\cite{Steiner13cs},
  SR1~\cite{Steiner05ia}, SR2~\cite{Steiner05ia},
  SR3~\cite{Steiner05ia}, es25~\cite{Steiner05ia},
  es275~\cite{Steiner05ia}, es30~\cite{Steiner05ia},
  es325~\cite{Steiner05ia}, es35~\cite{Steiner05ia},
  RAPR~\cite{Steiner05ia}, NL3~\cite{Lalazissis97},
  Z271~\cite{Horowitz01}, S271~~\cite{Horowitz01},
  NL4~\cite{NerloPomorska03}, FSUGold~\cite{ToddRutel05}, and
  IUFSU~\cite{Fattoyev10} models. We include several models from
  Ref.~\cite{Steiner05ia} because this paper contains several models
  with the same set of coupling constants. For each coupling (and also
  for the mass of the scalar meson, $m_{\sigma}$), for the results
  with and without Vela, we compute the quantity
\begin{equation}
  X_q \equiv \frac{P_q - \mathrm{min}_q}
  {\mathrm{max}_q- \mathrm{min}_q}
  \label{eq:X}
\end{equation}
where $P_q$ is the peak of the posterior distribution for quantity
$q$, and $\mathrm{min}_q$ and $\mathrm{max}_q$ are the minimum and
maximum values across the models listed above. The posterior value is
outside the range covered by the models only when $X<0$ or $X>1$. We
find values only slightly outside $[0,1]$: $a_3$ and $m_{\sigma}$ when
Vela is included (see Table I) and for $a_3$ and $a_4$ when Vela is
not included.

\begin{table}
  \begin{tabular}{lcc}
    quantity, $q$ & $X_{q,\mathrm{w/Vela}}$ & $X_{q,\mathrm{w/o~Vela}}$ \\
    \hline
    $a_1$ & 0.570 & 0.576 \\
    $a_2$ & 0.0168 & $-$5.91 $\times 10^{-3}$ \\
    $a_3$ & $-$0.112 & $-$0.0551 \\
    $a_4$ & 3.87 $\times 10^{-3}$ & $-$0.0545 \\
    $a_5$ & 8.61 $\times 10^{-4}$ & 0.0318 \\
    $a_6$ & 0.0697 & 0.0792 \\
    $b$ & 0.0787 & 0.124 \\
    $b_1$ & 0.899 & 0.767 \\
    $b_2$ & 1.01 & 0.988 \\
    $b_3$ & $-$3.33 $\times 10^{-4}$ & 1.37 $\times 10^{-4}$ \\
    $c$ & 0.248 & 0.196 \\
    $c_{\rho}$ & 0.977 & 0.937 \\
    $c_{\sigma}$ & 0.692 & 0.783 \\
    $c_{\omega}$ & 0.718 & 0.777 \\
    $m_{\sigma}$ & 1.25 & 0.876 \\
    $\xi$ & 6.18 $\times 10^{-3}$ & $-$5.50 $\times 10^{-3}$ \\
    $\zeta$ & 0.499 & 0.362 \\    
    \hline
  \end{tabular}
  \caption{Comparison of posteriors for the parameters in our 
    Lagrangian (given in Ref.~\cite{Steiner05ia})
    with the ranges given by other models (see Eq.~\ref{eq:X}).}
\end{table}

These deviations are relatively small, but it is more
  straightforward to interpret physical observables. We find the
  saturation density of nuclear matter, $0.155 \pm
  0.005~\mathrm{fm}^{-3}$ (with Vela) and $0.152 \pm
  0.004~\mathrm{fm}^{-3}$ (without Vela) are both in the empirical
  range~\cite{McDonnell15uq}. The binding energy at saturation is
  $-16.32 \pm 0.31~\mathrm{MeV}$ (with Vela) and $-16.34 \pm
  0.32~\mathrm{MeV}$ are in the empirical range as well.

\begin{table}
  \begin{tabular}{lrrrr}
    quantity & $C(q,K)_{\mathrm{w/Vela}}$ &
    $C(q,K)_{\mathrm{w/o Vela}}$ \\
    \hline
    $T_{C,n}$ & $-$0.0283 & 0.15 \\
    $k_{F,\mathrm{peak},n}$ & $-$0.143 & 0.596 \\
    $\Delta k_{F,n}$ & $-$0.179 & 0.534 \\
    $T_{C,p}$ & $-$0.0765 & 0.0189 \\
    $k_{F,\mathrm{peak},p}$ & $-$0.451 & $-$0.11 \\
    $\Delta k_{F,p}$ & 0.133 & 0.155 \\
    $x_{n=n_0}$ & 0.223 & 0.429 \\
    $x_{n=2 n_0}$ & 0.206 & 0.057 \\
    $x_{n=3 n_0}$ & 0.158 & $-$0.148 \\
    $M_{\mathrm{max}}$ & 0.912 & 0.543 \\
  \end{tabular}
  \caption{Pearson correlation coefficients for various quantities and
    the polytrope coefficient. The first six quantities are the
    parameters for the critical temperatures described in the method
    section above, $x$ is the proton fraction at the specified
    density, and the last row shows the correlation with the neutron
    star maximum mass.}
\end{table}

The additional polytrope, $P=K~\varepsilon^{\Gamma}$, is more
  unusual, so we estimate the impact that this additional polytrope
  has on the principal results of the paper. The posterior
  distribution for $K$ is $0.134 \pm 0.045$ (with Vela) and $0.100 \pm
  0.047$ (without Vela). The Pearson correlation coefficient between
  $K$ and several quantities from the simulation are given in Table
  II. When Vela is included, decreasing $K$ to zero (and thus removing
  the polytrope) will tend to decrease $k_{F,\mathrm{peak},p}$, and
  increase the proton fraction which may allow the direct Urca process
  to proceed in more massive stars. When Vela is not included,
  decreasing $K$ to zero will tend to increase both
  $k_{F,\mathrm{peak},n}$ and $\Delta k_{F,n}$ which will not have a
  strong impact. Since the polytrope modifies the high-density
  behavior, a correlation with the neutron star maximum mass,
  $M_{\mathrm{max}}$ is expected. Thus, removing the polytrope would
  modify the couplings of nuclear interaction, but those modifications
  are limited because of the constraints that we used. 


\end{document}